\documentclass[3p, showpacs, preprintnumbers, amsmath, amssymb]{elsarticle}

\usepackage{bm, amsfonts, mathtools}
\usepackage{graphicx, color, wasysym}
\usepackage{textcomp}
\usepackage{amsmath, amssymb, amsthm, latexsym, epsfig}
\usepackage{appendix}

\usepackage{stackrel}

\usepackage{txfonts}

\DeclareMathOperator{\D}{d\!}
\DeclareMathOperator{\E}{e} 
\DeclareMathOperator{\I}{i}

\theoremstyle{plain}
\newtheorem{thm}{Theorem}
\newtheorem{pro}[thm]{Proposition}
\newtheorem{lem}[thm]{Lemma}

\newtheorem{fact}[thm]{Fact}

\newtheorem{as}[thm]{Assumption}

\theoremstyle{definition}
\newtheorem{dfn}[thm]{{\it Definition}}
\newtheorem{exa}[thm]{{\it Example}}

\theoremstyle{remark}

\newtheorem*{prf}{{\it Proof}}

\begin{document} 

\allowdisplaybreaks

\makeatletter

\title{Non-Debye relaxations: two types of memories and their Stieltjes character}

\author[rvt]{K. G\'{o}rska} \corref{cor1}
\address[rvt]{Institute of Nuclear Physics, Polish Academy of Science, \\ ul. Radzikowskiego 152, 
PL-31342 Krak\'{o}w, Poland}
\cortext[cor1]{Corresponding author.}
\ead{katarzyna.gorska@ifj.edu.pl}
\author[rvt]{A. Horzela}
\ead{andrzej.horzela@ifj.edu.pl}

\begin{abstract}
We show that spectral functions relevant for commonly used models of the non-Debye relaxation are related to the  Stieltjes functions supported on the positive semiaxis. Using only this property it can be shown that the response and relaxation functions are nonnegative. They are connected to each other and obey the time evolution provided by integral equations involving the memory function $M(t)$ which is the Stieltjes function as well. This fact is also due to the Stieltjes character of the spectral function. Stochastic processes based approach to the relaxation phenomena  gives possibility to identify the memory function $M(t)$ with the Laplace (L\'evy) exponent of some infinitely divisible stochastic process and to introduce its partner memory $k(t)$. Both memories are related by the Sonine equation and lead to equivalent evolution equations which may be freely interchanged in dependence of our knowledge on memories governing the process.
\end{abstract}


\maketitle

\section{Introduction}\label{sec1}

The main information on the nature of dielectric relaxation phenomena comes from broadband dielectric spectroscopy  \cite{KremerSchoenhals} which provides us with data concerning dispersive and absorptive properties of dielectric materials. These properties are encoded in the complex dielectric permittivity $\hat{\varepsilon}(\I\!\omega)$ and its dependence on the frequency of external fields. In particular data of the spectroscopy experiments enable one to determine the behavior of permittivity $\hat{\varepsilon}(\I\!\omega)$ for asymptotic values of the frequency $\omega$ scaled with respect to some relaxation time $\tau$ which characterizes the material under investigation. Pioneer of this research was A. K. Jonscher who in 60's and 70's of the previous century studied, with his collaborators, a vast majority of available data and observed that they share universal asymptotic properties
\begin{align}\label{13/11-1}
\begin{split}
\hat{\varepsilon}(\I\!\omega) & \propto (\I\!\omega\tau)^{a-1}, \quad \omega\tau \ll 1 \\
  \Delta\hat{\varepsilon}(\I\!\omega) = \varepsilon_{0} - \hat{\varepsilon}(\I\!\omega) & 
\propto (\I\!\omega\tau)^{b}, \qquad \omega\tau \gg 1;
\end{split}
\end{align}
nowadays known as the Universal Relaxation Law (URL) \cite{AJonscher92}. In the above the static permittivity $\varepsilon_{0}$ denotes the limit of $\hat{\varepsilon}(\I\!\omega)$ for $\omega\to 0$ and the parameters $1-a$ and $b$ belong to the range $(0, 1)$. The Jonscher's URL agrees with the most commonly used phenomenological models  of the relaxation phenomena, namely with the Havrilak-Negami (HN) and the Jurlewicz-Weron-Stanislavski (JWS) models, both depending on a single characteristic time. For the HN model $b \geq a - 1$ which means that the exponent governing  asymptotics at infinity is larger than  its counterpart governing asymptotics at zero. The opposite situation occurs for the JWS model for which $b < a - 1$.

Recall that the normalized ratio of permittivities $[\hat{\varepsilon}(\I\!\omega) - \varepsilon_{\infty}]/[\varepsilon_{0} -\varepsilon_{\infty}]$ (here $\varepsilon_{\infty}$ means the infinite frequency limit of $\hat{\varepsilon}(\I\!\omega)$) is named the spectral function $\hat{\phi}(\I\!\omega)$. Through the Laplace transform it is related to the time domain response and relaxation functions, denoted as $\phi(t)$ and $n(t)$, respectively, 
\begin{equation}\label{13/11-10}
\hat{\phi}(\I\!\omega) = \mathcal{L}[\phi(t); \I\!\omega] \quad \text{and} \quad [1 - \hat{\phi}(\I\!\omega)]/(\I\!\omega) = \mathcal{L}[n(t); \I\!\omega], 
\end{equation}
which simply correspond to each other 
\begin{equation}\label{2/12-1}
\phi(t) = - \dot{n}(t). 
\end{equation}
{Actually, the transform in Eq. \eqref{13/11-10} is the Fourier transform restricted to the semiaxis but following physicists' customs we call it the Laplace transform \cite{CJFBoettcher96}.) Nevertheless, except the Subsec. 3.1 of the manuscript, we do not need to use the complex variables will show that it is enough to make considerations for functions supported  on the positive semiaxis. This is the reason why we will restrict ourselves  to the Laplace integral $\hat{f}(s) = \int_{0}^{\infty}\E^{-s t} f(t) \D t$ for $s > 0$. Its difference with respect to the Laplace transform $\hat{f}(z) = \int_{0}^{\infty}\E^{-z t} f(t) \D t$ is that $s$ is real while $z$ is complex. For $s > 0$ and $z\in \mathbb{C}\setminus\mathbb{R}^{-}$ they can be linked to each other through \cite[Theorem 2.6]{GGripenberg90}, repeatedly quoted in \cite[Theorem 1]{EDeOliveira11}. Because of the latter we feel free to use the same notation $\hat{f}$ for real and complex functions as well as to call interchangeably the $s$ or Laplace domain.}

The time evolution of the response function $\phi(t)$, {for physical reasons requriments to be continuous and vanishing for $t<0$}, is governed by the equation 
\begin{equation}\label{23/10-2}
\phi(t) = B M(t) - B \int_{0}^{t} M(t - \xi) \phi(\xi) \D\xi,
\end{equation}
where $B$ is a nonnegative transition rate constant and $M(t)$ is an integral kernel which plays the role of memory. {For consistency of its further physical and mathematical interpretation we assume that $M\in L^{1}_{\rm loc}(\mathbb{R}^{+})$.} {Due to the conditions given by \cite{AHanyga20, Kochubei11} there exists another function $k(t)\in L^{1}_{\rm loc}(\mathbb{R}^{+})$ which together with $M(t)$ satisfies the Sonine equation \cite{Sonine,  Kukushkin}
\begin{equation}\label{27/10-3}
\int_{0}^{t} k(u) M(t - u) {\rm d}u = \int_{0}^{t} k(t - u)M(u){\rm d}u = 1.
\end{equation}
In addition, assuming the conditions (*) in \cite{AHanyga20, Kochubei11} be fulfilled we can find that 
\begin{equation}\label{27/10-2}
\hat{k}(s) \hat{M}(s) = s^{-1}.
\end{equation}
}
{The authors of Refs. \cite{KGorska20, TSandev18} proposed to employ Eq. \eqref{27/10-2} to justify using an integro-differential equation}
\begin{equation}\label{26/10-1}
\frac{\D}{\D t} \int_{0}^{t} k(t - \xi) \phi(\xi) \D\xi = - B \phi(t)
\end{equation} 
as governing the time evolution of the response function. We emphasize that mathematical structure of the Eq. \eqref{26/10-1} was the subject of the seminal paper \cite{Kochubei11} whose results gave the conditions under which the Cauchy problem for the Eq. \eqref{26/10-1} is uniquely solved.

As it was shown in \cite{AStanislavsky20, AStanislavsky15, AStanislavsky19}, $\hat{M}(s)$ can be expressed by the algebraic inverse of the Laplace (L\'{e}vy) exponent $\Psi(s)$, i.e., $\hat{M}(s) = [\Psi(s)]^{-1}$,  of some infinitely divisible stochastic process $U$ underlying the relaxation. Mathematically, the Laplace (L\'{e}vy) exponent corresponds to the characteristic function of the process $U$ \cite{AStanislavsky20, AStanislavsky15}. For instance, the characteristic function of the L\'{e}vy stable process $X$ is given by the L\'{e}vy-Khintchine formula which naturally introduces $\Psi(s)$ as a complete Bernstein function (CBF).
\begin{dfn}
A real function $(0,\infty)\rightarrow h(s)$ is CBF if it is nonnegative infinitely differentiable function $h(s)$ which satisfies 
\begin{equation*}
(-1)^{n-1}h^{(n)}(s)\ge 0 \quad \text{for} \qquad n=1, 2, 3 \ldots
\end{equation*} 
and $h(s)/s$ is the Laplace transform (restricted to the positive semiaxis) of a CMF - the latter transform is equivalent to the Stieltjes transform (taken for the positive argument) of a nonnegative function supported on the positive semiaxis.
\end{dfn}
\noindent
{Thus, Eq. \eqref{27/10-2} illustrates relation between CBF and Sonine pair noticed only a few years ago \cite{Kochubei11}.}

{Completely Bernstein character of $\Psi(s)$} leads to the crucial observation concerning $\hat{M}(s)$ - because the algebraic inverse of a CBF is a Stieltjes function (SF) then  $\hat{M}(s)$ does share this property. {Following \cite{RLSchilling12, Kochubei11} we introduce 
\begin{dfn}\label{d25/01-1}
A (non-negative) SF is a function $f: (0, \infty)\mapsto [0, \infty)$ which can be written in the form 
\begin{equation}\label{25/01-1}
f(s) = a/s + b + \int_{0}^{\infty} \sigma(\D u)/(s + u),
\end{equation}
where $a, b \geq 0$ and $\sigma$ is a measure $(0, \infty)$ such that $\int_{0}^{\infty} \sigma(\D u)/(1 + u) < \infty$. 
\end{dfn} 
\noindent
We remark that if $a=0$ then the definition \ref{d25/01-1} is the same as given in \cite{Berg} and  \cite{NIAkhiezer65} but for both cases considered for complex numbers and a complex Stieltjes functions. Thus, SF coming from the definition \ref{d25/01-1} is restriction of a complex SF to the positive semiaxis. As an alternative definition of SF, justified for its convenience for futher considerations, we will use also Theorem 7.3 of \cite{RLSchilling12} which say that $f:(0, \infty) \mapsto [0, \infty)$ is SF if, and only if, $1/f$ is CBF. } Note that SFs form a subclass of completely monotonic functions (CMF) defined as. 
\begin{dfn}
A real function $\hat{c}(s): (0, \infty) \mapsto \mathbb{R}$ is CMF if all its derivatives for $n = 0, 1, 2, \ldots$ derivatives and $(-1)^{n} \hat{c}^{(n)}(s) \geq 0$.
\end{dfn}
\noindent
The above implies that $\hat{M}(s)$, as a reciprocal of a CBF, is CMF as well and, according to Refs.  \cite{RSAnderssen02, RSAnderssen02a}, we can call such completely monotonic integral kernels as fading memories. 

Physical interpretation of Eqs. \eqref{23/10-2} and \eqref{26/10-1} yields that they should lead to the same physical results. To endowe this property mathematical meaning note that the first of them is the integral equation whereas the next one is an integro-differential equation. Recall that in the theory of integral equations the homogeneous integral equation can be transformed to its  differential analogue - in what follows we are going to demonstrate that analogical procedure may be performed for Eqs. \eqref{23/10-2} and \eqref{26/10-1}. It means that from Eq. \eqref{23/10-2} we should derive Eq. \eqref{26/10-1}. Doing that we will see the importance of the condition \eqref{27/10-3} or its analogue in the Laplace domain Eq. \eqref{27/10-2}. Moreover, to name the integral kernels $M(t)$ and $k(t)$ the fading memories it will be essential to prove that they are given by SFs being the subclass of CMFs. In the paper  we will show all these properties using the fact that the spectral function in the complex domain can be rewritten as SF. 

Our presentation goes as follows. Sec. \ref{sec2} shows that for commonly used models of non-Debye relaxations their spectral functions are given by SFs. Thus it appears that the response and relaxation functions are nonnegative. In Sec. \ref{sec3} we show that assuming the Stieltjes character of the spectral functions is enough to obtain two types of integral kernels which govern the evolution of the response functions and which also are SFs. Hence, we can call them the memory functions. We will give the exact and explicite forms of these memories. Requiring that both equations give the same result we conclude that the memories have to satisfy the Sonine equation. The paper is summarized in Sec. \ref{sec5}.

\section{Basic models of non-Debye relaxations}\label{sec2}

The spectral functions of HN and JWS models, in physical literature called also relaxation patterns, from construction depend on a single characteristic time $\tau$ and are given by
\begin{equation}\label{5/11-1}
\hat{\phi}_{HN; \,\alpha, \beta}(\I\!\omega) = [1 + (\I\!\omega\tau)^{\alpha}]^{-\beta} \quad \text{and} \quad \hat{\phi}_{JWS; \,\alpha, \beta} (\I\!\omega) = 1 - (\I\!\omega\tau)^{\alpha\beta}\,\hat{\phi}_{HN; \,\alpha, \beta}(\I\!\omega),
\end{equation}
where $\alpha, \beta \in [0, 1]$. Values of  $\alpha$ and $\beta$ are obtained from the experiment so they correspond to the URL. For the HN relaxation we have $a = 1 - \alpha\beta$ and $b = \alpha$ whereas for the JWS model we get $a = 1 - \alpha$ and $b = \alpha\beta$. For special choices of the parameters the HN and JWS  patterns boil down to other widely used models of non-Debye relaxations namely the Cole-Davidson (CD), the Cole-Cole (CC), and the Debye relaxation (D). 
The CD model is obtained from the HN model for $\alpha = 1$ and $\beta\in[0, 1]$ whereas the Cole-Cole pattern (CC) is got either from the HN or from the JWS model for $\alpha\in[0, 1]$ and $\beta = 1$. The HN and JWS models for $\alpha = \beta = 1$ reduce to the Debye relaxation (D). If the frequency of applied electric field is of the order $10^{5} - 10^{10}$ Hz then fitting the experimental data by a single standard relaxation pattern is not satisfactory any longer and it is much more effective to fit them by a (linear) combination of above mentioned non-Debye relaxation models or to use models which belong to the excess wing model class (EW). The latter involve the extended number of parameters, in particular introduce more than one characteristic time \cite{RHilfer17}. Thus, in the high frequency domain we deal with more than one characteristic time scale and the Jonscher's URL is not satisfied any longer. For instance, in the simplest version of the EW model we have two characteristic times, $\tau_{1}$ and $\tau_{2}$, built into the spectral function as follows:
\begin{equation}\label{5/11-2}
\hat{\phi}_{EW; \, \alpha}(\I\!\omega) = \frac{1 + (\I\!\omega\tau_{2})^{\alpha}}{1 + \I\!\omega\tau_{1} + (\I\!\omega\tau_{2})^{\alpha}}, \qquad 0 \leq \alpha \leq 1.
\end{equation}
The HN, JWS, and EW spectral functions, in common denoted as $\hat{\phi}_{(\cdot)}(\I\!\omega)$ are the (complex) Stieltjes functions which form the subclass of the Nevanlinna-Pick functions $P(z)$, analytic in the upper half plane and satisfying ${\rm Im}P(z)\ge 0$ for ${\rm Im}\, z > 0$ \cite{Berg, NIAkhiezer65, RLSchilling12}. Analyticity in the upper half plane guarantees that the Kramers-Kronig relations are satisfied \cite{CJFBoettcher96}. The link between the Nevanlinna-Pick functions and the CMFs is presented by \cite[Theorem 2.6]{GGripenberg90}, repeatedly quoted in \cite[Theorem 1]{EDeOliveira11}. As is signalized in Sec. \ref{sec1} we will provide our studies in the real domain.

With the help of just mentioned theorem Eqs. \eqref{5/11-1} and \eqref{5/11-2} can be rewritten as
\begin{equation}\label{5/11-3}
\hat{\phi}_{HN; \,\alpha, \beta}(s) = [1 + (s \tau)^{\alpha}]^{-\beta} \quad \text{and} \quad\hat{\phi}_{JWS; \,\alpha, \beta}(s) = 1 - (s\tau)^{\alpha\beta}\,\hat{\phi}_{HN; \alpha, \beta}(s)
\end{equation}
while
\begin{equation}\label{5/11-4}
\hat{\phi}_{EW; \alpha}(s) = \frac{1 + (s\tau_{2})^{\alpha}}{1 + s\tau_{1} + (s\tau_{2})^{\alpha}} =\frac{1}{1 + \tau_{1} s/[1 + (s \tau_2)^{\alpha}]},
\end{equation}
where $s>0$ and $0 \leq \alpha, \beta \leq 1$. 

All these functions are SFs and their Stieltjes character can be shown using the power function $s^{\mu}$ which due to the value of exponent $\mu$ is either CMF, or SF, or CBF. Namely, for $\mu \leq 0$ it is CMF, for $\mu\in[-1, 0]$ it is SF, and for $\mu\in[0, 1]$ it is CBF. Let us now check if the considered spectral functions really belong to SFs.
\begin{fact}\label{5/11-1f}
{\rm From property (cb1) it appears that the convex sum $1 + \tau^{\alpha} s^{\alpha}$, $\alpha\in[0,1]$ is  CBF. Moreover, property (cb2)  gives that the composition of SF (here $\sigma^{-1}$ with $\sigma > 0$) and CBF (here $1 + \tau^{\alpha} s^{\alpha}$) is SF. Thus, $\hat{\phi}_{HN; \,\alpha, \beta}(s)$ is SF. \\
The HN spectral function is bounded. Its maximal value equals $1$ and the function $\hat{\phi}_{HN; \,\alpha, \beta}$ decreases to $0$ at infinity.} 
\end{fact}
\begin{fact}\label{5/11-2f}
{\rm Because $f(s) = [1 + (s\tau)^{\alpha}]^{\beta}$ is CBF for $ 0 \leq \alpha, \beta \leq 1$ then the property (cb3) implies that 
\begin{equation*}
\frac{1}{f(1/s)} = \frac{1}{[1 + (s \tau)^{-\alpha}]^{\beta}} = \frac{(s\tau)^{\alpha\beta}}{[1 + (s\tau)^{\alpha}]^{\beta}} = (s\tau)^{\alpha\beta}\,\hat{\phi}_{HN; \,\alpha, \beta}(s) 
\end{equation*}
is CBF because it is linear combination of CBFs with all weight functions equal to 1. The pointwise limit of this combination gives the series for $(s\tau)^{\alpha\beta} \hat{\phi}_{HN; \,\alpha\beta}(s)$. Then, 
\begin{equation}\label{22/11-1}
\sum_{r=0}^{\infty} [(s\tau)^{\alpha\beta} \hat{\phi}_{HN; \,\alpha\beta}(s)]^{r} = [1 - (s\tau)^{\alpha\beta} \hat{\phi}_{HN; \,\alpha\beta}(s)]^{-1},
\end{equation}
{The nonnegative function $(s\tau)^{\alpha\beta}\,\hat{\phi}_{HN; \,\alpha, \beta}(s)$ varies from 0 to 1 with increasing $s > 0$. Thus, we do not need any further restrictions put on the elements of series in Eq. \eqref{22/11-1} - the equality in Eq. \eqref{22/11-1} is universally true.} With a little help of the property (cb7) we obtain that Eq. \eqref{22/11-1} is CBF. Thus, using the definition of SF given by \cite[Theorem 7.3]{RLSchilling12} or property (cb2) gives that $\hat{\phi}_{JWS; \, \alpha, \beta}(s)$ is SF. \\
The JWS spectral function for $s > 0$ decreases from $\hat{\phi}_{JWS; \,\alpha, \beta}(0)=1$ (which is its maximal value) to $0 = \lim\limits_{s\to\infty}\hat{\phi}_{JWS;\, \alpha, \beta}(s)$ being the minimal value of $\hat{\phi}_{JWS;\, \alpha, \beta}(s)$.
}
\end{fact}
\begin{fact}\label{5/11-3f}
{\rm Because $1 + (s \tau_{2})^{\alpha}$ is CFB for $0 \leq \alpha \leq 1$ then from property (cb5) it appears that $\tau_{1} s/[1 + (s \tau_{2})^{\alpha}]$ for $\tau_{1} > 0$ is also CBF. From property (cb8) we conclude that the EW spectral function $\{1 + \tau_{1} s/[1 + (s \tau_{2})^{\alpha}]\}^{-1}$ is SF which decreases from $1$ to $0$ with increasing $s$.}
\end{fact}
\begin{as}\label{a1}
The spectral function $\hat{\phi}_{(\cdot)}(s)$ for $s > 0$ is given by a bounded SF.
\end{as}

At this point of our considerations we make a comment concerning physical meaning of our work. Studying  measured experimental data we are able to determine only the relaxation function, its first time derivative and eventually the second one. Therefore the requirement that all derivatives exist and alternate for $s>0$, as it is needed by the definition of CMF, is impossible to be verified in practice. Nevertheless, from just listed Facts \ref{5/11-1f}-\ref{5/11-3f} we learn that the spectral functions used to fit the data are modelled by SFs, and thus CMFs. Moreover, Assumption \ref{a1} simplify many calculations because we know that any CMF is uniquely represented by a Laplace integral of a nonnegative function. {Indeed, the Bernstein theorem (called also the Bernstein-Widder theorem) according to the classical D. V. Widder's book reads \cite[Theorem 12a]{DVWidder}:
\begin{thm}
A necessary and sufficient condition that $f(x)$ should be completely monotonic in $0 \leq x < \infty$ is that
\begin{equation}\label{25/01-2}
f(x) = \int_{0}^{\infty} \E^{-x \xi} \D\alpha(\xi),
\end{equation}
where $\alpha(\xi)$ is bounded and non-decreasing and the integral converges for $0 \leq x < \infty$.
\end{thm}
\noindent
The proof of the Bernstein theorem can be found in \cite{DVWidder, HPollard44}. We notice that the formulation of the Bernstein theorem can confuse the reader because some of the authors use the Laplace transform (e.g. \cite{Kochubei11}) but in Widder's and Pollard's approach  the theorem is formulated  just in terms of the real valued integral. For majority of physicists the name "integral transform"  means that we can invert Eq. \eqref{25/01-2} which usually demands using methods of the complex analysis. Here, such considerations are not needed. Hence we will use Widder's and his student, H. Pollard, definition \cite{DVWidder, HPollard44}.} Thanks to the Bernstein theorem we can claim that 
\begin{pro}\label{cor1}
The response function $\phi(t)$ and the relaxation function $n(t)$ are nonnegative.
\end{pro}
\begin{prf}
The proof of nonnegativity of $\phi(t)$ flows immediately from the Bernstein theorem applied to the first formula of Eqs. \eqref{13/11-10} which written in the $s$ domain reads
\begin{equation}\label{13/11-11}
\hat{\phi}(s) = \int_{0}^{\infty} \E^{-st} \phi(t) \D t, \qquad \text{where} \quad \phi(t) \geq 0.
\end{equation}
To show that the relaxation function $n(t)$ is given by a nonnegative function we use the second formula of Eqs. \eqref{13/11-10} in the $s$ domain, this is
\begin{equation}\label{2/11-1}
[1 - \hat{\phi}(s)]/s = \int_{0}^{\infty} \E^{- s t} n(t) \D t. 
\end{equation}
Note that the function $1 - \hat{\phi}(s)$ can be rewritten as $\hat{\phi}(0) - \hat{\phi}(s)$ where {$\hat{\phi}(0)$ is bounded and} $\hat{\phi}(0) = 1$ is the maximum of $\hat{\phi}(s)$. Then, the property (cb4) implies that it is CBF and, then, from definition of CBF it appears that $[1 - \hat{\phi}(s)]/s$ is SF (property (cb6)). Furthermore, Eq. \eqref{2/11-1} means that the Laplace integral of $n(t)$ is equal to $[1 - \hat{\phi}(s)]/s$ which is SF and thus CMF. Then, the Bernstein theorem implies that $n(t)$ is nonnegative. \qed
\end{prf}

The exact forms of the response function $\phi(t)$ and the relaxation function $n(t)$ for the HN and JWS models can be found in, e.g., \cite{RGarrappa16, KGorska18}. Relevant formulae are
\begin{equation}\label{14/11-1}
n_{HN;\, \alpha, \beta}(t) = 1 - (t/\tau)^{\alpha\beta} E_{\alpha, 1 + \alpha\beta}^{\beta}[-(t/\tau)^{\alpha}],  \qquad  \phi_{HN;\, \alpha, \beta}(t) = \tau^{-1} (t/\tau)^{\alpha\beta - 1} E_{\alpha, \alpha\beta}^{\beta}[-(t/\tau)^{\alpha}],
\end{equation}
and
\begin{equation}\label{8/11-15}
n_{JWS;\, \alpha, \beta}(t) = E_{\alpha, 1}^{\beta}[-(t/\tau)^{\alpha}], \qquad \phi_{JWS;\, \alpha, \beta}(t) = \delta(t) - \tau^{-1}(t/\tau)^{-1} E_{\alpha, 0}^{\beta}[-(t/\tau)^{\alpha}]. 
\end{equation}
Using the response and relaxation functions for the HN model we can check the correctness of Eq. \eqref{2/12-1} - it immediately flows out from Eq. \eqref{15/11-2b}. From Eq. \eqref{2/12-1} and Eqs. \eqref{8/11-15} one finds the formula which describes the first derivative of $E_{\nu, \mu}^{\lambda}(a x^{\nu})$. Observe, that in this case Eq. \eqref{15/11-2b} cannot be used for the JWS relaxation because it works only if the order of derivative is larger than the value of second lower parameter in the Mittag-Leffler function $E_{\nu, \mu}^{\lambda}(a x^{\nu})$. Eqs. \eqref{2/12-1} and  \eqref{8/11-15} give
\begin{equation}\label{16/11-1}
\frac{\D}{\D x} E_{\nu, 1}^{\lambda}(a x^{\nu}) = x^{-1} E_{\nu, 0}^{\lambda}(a x^{\nu}) - \delta(x), \qquad x\in\mathbb{R}.
\end{equation}

For the EW model  only the relaxation function $n_{EW;\, \alpha}(t)$ is known.  It can be expressed through the binomial Mittag-Leffler function \cite{RHilfer17}
\begin{equation}\label{8/11-16}
n_{EW;\, \alpha}(t) = E_{(1, 1 - \alpha), 1}(-t/\tau_{1}, - \tau_{2}^{\alpha}\, t^{1-\alpha}/\tau_{1}) 
\end{equation}
or as the series of three parameter Mittag-Leffler function \cite[Eqs. (3.71) or (3.73)]{RGarrappa16}. These series lead to Eq. \eqref{8/11-16} with the help of Eqs. \eqref{23/11-1}. To calculate the response function $\phi_{EW;\, \alpha}(t)$ we make the inverse Laplace transform of the spectral function Eq. \eqref{5/11-4} which can be easily get with the help of Eq. \eqref{15/11-1b}. That enables us to write 
\begin{equation}\label{15/11-1}
\phi_{EW;\, \alpha}(t) = \delta(t) - t^{-1} E_{(1, 1-\alpha), 0}(-t/\tau_{1}, - \tau_{2}^{\alpha}\, t^{1-\alpha}/\tau_{1}).
\end{equation}
Analogically as in the derivation of Eq. \eqref{16/11-1} we can obtain the first derivative of the binomial Mittag-Leffler function $E_{(\nu_{1}, \,\nu_{2}), \,1}(a x^{\nu_{1}}, \,b x^{\nu_{2}})$. It reads
\begin{equation}\label{23/11-2}
\frac{\D}{\D x} E_{(\nu_{1},\, \nu_{2}),\, 1}(a x^{\nu_{1}}, b x^{\nu_{2}}) = x^{-1} E_{(\nu_{1},\, \nu_{2}), \,0}(a x^{\nu_{1}}, b x^{\nu_{2}}) - \delta(x), \qquad x\in\mathbb{R}.
\end{equation}
Eq. \eqref{23/11-2} can be also derived by employing the representation of the binomial Mittag-Leffler function in terms of the series of three parameter Mittag-Leffler functions given by Eq. \eqref{23/11-1} and, next, separate from these series the zero term, i.e. $E_{\nu_{1}, 1}^{1}(b x^{\nu_{2}})$ or $E_{\nu_{2}, 1}^{1}(a x^{\nu_{1}})$. The calculation goes as follows
\begin{align*}
\frac{\D}{\D x} E_{(\nu_{1}, \nu_{2}), 1}(a x^{\nu_{1}}, b x^{\nu_{2}}) & = \frac{\D}{\D x} \sum_{r \geq 0} (a x^{\nu_{1}})^{r} E_{\nu_{2}, \nu_{1} r + 1}^{1+r}(b x^{\nu_{2}}) \\
& = \frac{\D}{\D x} E_{\nu_{2}, 1}^{1}(b x^{\nu_{2}}) + \sum_{r \geq 1} a^{r} \frac{\D}{\D x} [x^{\nu_{1} r} E_{\nu_{2}, \nu_{1} r + 1}^{1+r}(b x^{\nu_{2}})].
\end{align*}
Now, applying Eq. \eqref{16/11-1} and Eq. \eqref{15/11-2b} we get
\begin{equation*}
\frac{\D}{\D x} E_{(\nu_{1}, \nu_{2}), 1}(a x^{\nu_{1}}, b x^{\nu_{2}}) = x^{-1} E_{\nu_{2}, 0}^{1}(b x^{\nu_{2}}) + x^{-1} \sum_{r \geq 1} (a x^{\nu_{2}})^{r} E_{\nu_{1}, \nu_{2}r}^{1 + r}(b x^{\nu_{2}}) - \delta(x),
\end{equation*}
which restores Eq. \eqref{23/11-2}. The same way can be repeated for the representation of $E_{(\nu_{1}, \nu_{2}), 1}(a x^{\nu_{1}}, b x^{\nu_{2}})$ through the second series in Eq. \eqref{23/11-1}.

The series form of the three parameter Mittag-Leffler function $E_{\nu, \mu}^{\lambda}(x)$ for $x\in\mathbb{R}$ and the binomial Mittag-Leffler function $E_{(\nu_{1}, \nu_{2}), \mu}(x, y)$ for $x, y\in\mathbb{R}$ as well as their properties are recalled in Appendix \ref{app2}.

\section{$\hat{M}(s)$ and $\hat{k}(s)$ and the Laplace (L\'{e}vy) exponents $\Psi(s)$ related to them}\label{sec3}

In this section we shall show that to determine the Stieltjes character of $\hat{M}(s)$ and $\hat{k}(s)$ as well as to explain the CBF nature of the Laplace (L\'{e}vy) exponent $\Psi(s)$ it is enough to demand the Assumption 1 be fulfilled.

Eqs. \eqref{23/10-2} and \eqref{2/11-1} allows one to express the memory $M(t)$ in terms of the spectral function $\hat{\phi}(s)$. From them we have
\begin{equation}\label{3/11-4}
\hat{M}(s) = B^{-1} \hat{\phi}(s) [1 - \hat{\phi}(s)]^{-1},
\end{equation}
which is SF.
\begin{prf}
The proof of the Stieltjes character of $\hat{M}(s)$ goes as follows. We begin with $\sum_{r=0}^{n}[\hat{\phi}(s)]^{r+1}$ which is SF as a linear combination of SFs. Then, taking the pointwise limit $n\to \infty$ we obtain the series $\sum_{r=0}^{\infty}[\hat{\phi}(s)]^{r+1}$ which {tends to $B \hat{M}(s)$ for all $\hat{\phi}(s)$ (we remind that $\hat{\phi}(s)$ is a bounded SF which maximum is $\hat{\phi}(0) = 1$ and which vanishes for $s\to\infty$).} Due to property (s1) we end up the proof. \qed
\end{prf}

The algebraic inverse of $\hat{M}(s)$ gives the Laplace (L\'{e}vy) exponent $\Psi(s)$:
\begin{equation}\label{4/11-1}
\Psi(s) = B [1 - \hat{\phi}(s)]/\hat{\phi}(s),
\end{equation}
which is CBF. The complete Bernstein character of $\Psi(s)$ is confirmed by the standard non-Debye relaxation, i.e. by the HN, JWS, and EW models. That is illustrated in the examples \ref{8/11-1e} - \ref{8/11-3e} below.
\begin{exa}\label{8/11-1e}
{ For the HN relaxation $\Psi_{HN; \, \alpha, \beta}(s)$ reads $B \{[1 + (\tau s)^{\alpha}]^{\beta} - 1\}$. Thus, $[1 + \Psi_{HN; \, \alpha, \beta}(s)/B]^{-1}$ gives $[1 + (\tau s)^{\alpha}]^{-\beta}$ which is SF for $\alpha, \beta\in [0, 1]$. The property (s2) and positivity of $B$ implies that $\Psi_{HN; \, \alpha, \beta}(s)$ is CBF.
}
\end{exa}
\begin{exa}\label{8/11-2e}
{ In the case of JWS model we express the Laplace (L\'{e}vy) exponent as $\Psi_{JWS; \, \alpha, \beta}(s) = [\Psi_{HN; \, \alpha, \beta}(1/s)]^{-1}$. From the property (cb3) it emerges that it is CBF.}
\end{exa}
\begin{exa}\label{8/11-3e}
{ The EW Laplace exponent reads $\Psi_{EW;\, \alpha}(s) = B s\tau_{1}/[1 + (\tau_{2} s)^{\alpha}]$ and it is CBF. $1 + (\tau_{2} s)^{\alpha}$ is CBF and grace to the property (cb5) $s/[1 + (\tau_{2} s)^{\alpha}]$ is also CBF.}
\end{exa}

The Laplace (L\'{e}vy) exponent $\Psi(s)$ can be also employed for calculating $\hat{k}(s)$ which through Eq. \eqref{27/10-2} are coupled to $\hat{M}(s)$. It reads
\begin{equation}\label{4/11-3}
\hat{k}(s) = \Psi(s)/s = B [1 - \hat{\phi}(s)]/[s \hat{\phi}(s)] \quad \text{and it is SF.}
\end{equation}
The Stieltjes character of $\hat{k}(s)$ flows out from the fact that $\Psi(s)$ is CBF and the property (cb6). Moreover, from the property (cb5) it occurs that $s/\Psi(s) = \Phi(s)$ is CBF such that it can be treated as another Laplace (L\'{e}vy) exponent.

\subsection{Examples of memories $M(t)$ and $k(t)$}

The examples of $\hat{M}(s)$ and $\hat{k}(s)$ in the $s$ domain for the HN, JWS, and EW models are listed in Table \ref{tab1},
\begin{table}[h]
\caption{\label{tab1} $\hat{M}(s)$ and $\hat{k}(s)$ for the HN, JWS, and EW models.}
\centering
\begin{tabular}{c | c | c}
\\
	& $\boldsymbol{\hat{M}(s)}$	& $\boldsymbol{\hat{k}(s)}$ \\ \hline
\\
HN     	&  $B^{-1}\{[1 + (\tau s)^{\alpha}]^{\beta} - 1\}^{-1}$  & $B s^{-1} [1 + (\tau s)^{\alpha}]^{\beta} - s^{-1}$ \\
\\
JWS		&  $ B^{-1}[1 + (\tau s)^{-\alpha}]^{\beta} - 1$        & $B s^{-1}\{[1 + (\tau s)^{-\alpha}]^{\beta} - 1\}^{-1}$\\
\\
EW   	&  $B^{-1}(\tau_{2}^{-\alpha} + s^{\alpha})/s$	      & $B(\tau_{2}^{-\alpha} + s^{\alpha})^{-1}$\\

\end{tabular}
\end{table}
\noindent
whereas in the time $t$ domain we have Table \ref{tab2}. 
\begin{table}[!h]
\caption{\label{tab2} $M(t)$ and $k(t)$ for the HN, JWS, and EW models.}
\centering
\begin{tabular}{c  | c | c}
\\
	& $\boldsymbol{M(t)}$	& $\boldsymbol{k(t)}$\\ \hline
\\
HN     	&  $(B t)^{-1} \sum_{r\geq 0} (t/\tau)^{\alpha\beta(r+1)} E_{\alpha, \alpha\beta (r + 1)}^{\beta (r + 1)}[-(t/\tau)^{\alpha}]$  & $B (\tau/t)^{\alpha\beta} E_{\alpha, 1-\alpha\beta}^{-\beta}[-(t/\tau)^{\alpha}] - B$ \\
\\
JWS		&  $(B t)^{-1} E_{\alpha, 0}^{-\beta}[-(t/\tau)^{\alpha}] - B^{-1}\delta(t)$  & $B \sum_{r\geq 0} E_{\alpha, 1}^{\beta(r+1)}[-(t/\tau)^{\alpha}]$\\
\\
EW   	&  $B^{-1} \tau_{2}^{-\alpha} + B^{-1}t^{-\alpha}/\Gamma(1-\alpha)$	      & $B t^{\alpha - 1} E_{\alpha, \alpha}[-(t/\tau_{2})^{\alpha}]$\\
\end{tabular}
\end{table}
\noindent
Relations  between functions listed in Tables \ref{tab1} and \ref{tab2} are obtained through the Laplace transform in which we use the complex $z$ instead of $s > 0$. 

\section{Equivalence of Eqs. \eqref{23/10-2} and \eqref{26/10-1}}\label{sec4}

\begin{lem}\label{27/01-1}
Eqs. \eqref{23/10-2} and \eqref{26/10-1} are equivalent.
\end{lem}
\begin{prf}
To show that {Lemma \ref{27/01-1} is true} we integrate both sides of Eq. \eqref{23/10-2} taking $\int_{0}^{T} k(T - t) \cdots \D t$ of it where instead of "$\cdots$" we substitute all terms of Eq. \eqref{23/10-2}. In this way we obtain
\begin{equation}\label{29/10-4}
\int_{0}^{T} k(T - t) \phi(t) \D t =  B \int_{0}^{T} k(T - t) M(t) \D t - B \int_{0}^{T} k(T - t) \left[\int_{0}^{t} M(t - \tau) \phi(\tau) \D\tau\right] \D t.
\end{equation}
Afterwards, due to the Dirichlet formula, the double integral $\int_{0}^{T} \D t \int_{0}^{t} \D\tau$ can be changed to $\int_{0}^{T} \D\tau \int_{\tau}^{T} \D t$. Consequently, Eq. \eqref{29/10-4} is expressed as
 \begin{equation}\label{29/10-2}
 \int_{0}^{T} k(T - t) \phi(t) {\rm d}t = B \int_{0}^{T} k(T - t) M(t) \D t - B \int_{0}^{T} \left[ \int_{\tau}^{T} k(T - t) M(t - \tau) \phi(\tau) \D t \right] \D\tau.
 \end{equation}
Setting $T - t = u$ we transform Eq. \eqref{29/10-2} as
\begin{align}\label{29/10-3}
\begin{split}
 \int_{0}^{T} k(T - t) \phi(t) \D t & = B \int_{0}^{T} k(T - t) M(t) \D t - B \int_{0}^{T} \left[\int_{0}^{T-\tau} k(u) M(T-\tau - u) \D u\right] \phi(\tau) \D\tau \\
 & = B - B \int_{0}^{T} \phi(\tau) \D\tau,
\end{split}
\end{align}
where the passage from upper to lower formulas goes by using Eq. \eqref{27/10-3} twice. Next, we differentiate it with respect to $T$. That ends the proof. \qed 
\end{prf}

\begin{exa}\label{19/11-1}
{\rm In the case of the HN model only the memory $k_{HN; \alpha, \beta}(t)$ is known explicitly. Thus Eq. \eqref{26/10-1} is much simpler to find the equation governing behavior of the HN response function. The substitution of $k_{HN; \alpha, \beta}(t)$ into Eq. \eqref{26/10-1} leads to \cite[Eq. (3.36)]{RGarrappa16}, namely
\begin{equation}\label{19/11-2}
({_{0}D_{t}^{\alpha}} + \tau^{-\alpha})^{\beta} \phi(t) = \frac{\D}{\D t} \int_{0}^{t} \Big(\frac{t-\xi}{\tau}\Big)^{-\alpha\beta} E_{\alpha, 1 - \alpha\beta}^{- \beta}\left[-\Big(\frac{t-\xi}{\tau}\Big)^{\alpha}\right] \phi(\xi) \D\xi = 0.
\end{equation}
The symbol ${_{0}D_{t}^{\alpha}}$ denotes the fractional derivative in the Riemann-Liouville sense. {The right (called also upper) fractional derivative in the Riemann-Liouville sense ${_{0}D_{t}^{\alpha}}$ can be defined through the fractional integral $({_{0}I_{t}^{\alpha}} f)(t) = ({_{0}D_{t}^{-\alpha}} f)(t) = [\Gamma(\alpha)]^{-1} \int_{0}^{t} (t - \xi)^{\alpha-1} f(\xi) \D \xi$ as follows $({_{0}D_{t}^{\alpha}} f)(t) = \D^{\,n}\! ({_{0}I_{t}^{n - \alpha}} f)(t)/ \D t^{\, n}$. Furthermore, the fractional derivative in the Caputo sense $({_{0}^{C}D_{t}^{\alpha}} f)(t)$ is equal to ${_{0}I_{t}^{n - \alpha}} \D^{\,n}\! f(t)/ \D t^{\, n}$.} The complete information how to understand the  pseudo-operator $(\,{_{0}D_{t}^{\alpha}} + \tau^{-\alpha})^{\beta}$ can be found in \cite[Appendix B]{RGarrappa16} and in {\cite[Section 5]{AGusti20}}.
}
\end{exa}
\begin{exa}\label{19/11-4}
{\rm For the JWS model the memory $M_{JWS; \,\alpha, \beta}(t)$ is exactly known. It makes possible to get Eq. \eqref{23/10-2} relevant for this model. To achieve this goal we represent $M_{JWS; \,\alpha, \beta}(t)$ taken from the Table \ref{tab2} as the fractional derivative in the Riemann-Liouville sense minus a term being the Dirac $\delta$ distribution, i.e. 
\begin{equation}\label{19/11-5}
M_{JWS; \,\alpha, \beta}(t) = {_{0}D_{t}^{1-\alpha\beta}}\{t^{-\alpha\beta} E_{\alpha, 1 - \alpha\beta}^{-\beta}[-(t/\tau)^{\alpha}]\} - \delta(t)
\end{equation}
and expressing $M_{JWS; \,\alpha, \beta}(t)$ with the help of Eq. \eqref{16/11-1} as the first derivative. Then, substituting it into Eq. \eqref{23/10-2} we get
\begin{equation}\label{19/11-6}
\int_{0}^{t} {_{0}D_{t}^{1-\alpha\beta}} \{(t-\xi)^{-\alpha\beta} E_{\alpha, 1 - \alpha\beta}^{-\beta}[-(t - \xi)^{\alpha}/\tau^{\alpha}]\phi(\xi) \} \D\xi = \frac{\D}{\D t} E_{\alpha, 1}^{-\beta}[-(t/\tau)^{\alpha}]
\end{equation}
Integrating both sides of this equation with ${_{0}D_{t}^{-(1-\alpha\beta)}}$ and using the fact that ${_{0}D_{t}^{-\mu}} [{_{0}D_{t}^{\mu}}f(t)]= f(t) - t^{\mu-1}/\Gamma(\mu) \times[{_{0}D_{t}^{\mu-1}} f(t)]_{t = 0}$ given in \cite[Eq. (2.113) on p. 70]{IPodlubny99} we obtain
\begin{equation}\label{19/11-7}
\int_{0}^{t} (t-\xi)^{-\alpha\beta} E_{\alpha, 1 - \alpha\beta}^{-\beta}[-(t-\xi)^{\alpha}/\tau^{\alpha}] \phi(\xi) \D\xi = t^{-\alpha\beta} E_{\alpha, 1 - \alpha\beta}^{-\beta}[-(t/\tau)^{\alpha}] - \frac{t^{-\alpha\beta}}{\Gamma(1 - \alpha\beta)}.
\end{equation}
This formula is the same as \cite[Eq. (3.36)]{RGarrappa16}.
}
\end{exa}
\begin{exa}\label{19/11-9}
{\rm In the case of EW relaxation $M_{EW; \alpha}(t)$ and $k_{EW; \alpha}(t)$ are known in compact forms so Eqs. \eqref{23/10-2} and \eqref{26/10-1} can be found either. After substituting $M_{EW; \alpha}(t)$ from Table \ref{tab2} Eq. \eqref{23/10-2} reads
\begin{equation}\label{15/12-1}
\phi_{EW; \alpha}(t) = \tau_{2}^{-\alpha}\Big[1 - \int_{0}^{t} \phi_{EW; \alpha}(\xi) \D\xi\Big] + \frac{t^{-\alpha}}{\Gamma(1-\alpha)} + {_{0}D_{t}^{-(1-\alpha)}} \phi_{EW; \alpha}(t),
\end{equation}
where ${_{0}D_{t}^{-(1-\alpha)}}$ is a fractional integral given below Eq. \eqref{19/11-2}. Using Eq. \eqref{2/12-1} we can find the evolution equation for $n_{EW; \alpha}(t)$. Namely, $1 - \int_{0}^{t} \phi_{EW; \alpha}(\xi) \D\xi = n_{EW; \alpha}(t)$ and ${_{0}D_{t}^{-(1-\alpha)}} \phi_{EW; \alpha}(t) = {_{0}^{C}D_{t}^{\,\alpha}}\,n_{EW; \alpha}(t)$. Moreover, $t^{-\alpha}/\Gamma(1-\alpha) + {_{0}^{C}D_{t}^{\,\alpha}}\,n_{EW; \alpha}(t)$ is equal to the fractional derivative of the Riemman-Liouville sense, i.e. ${_{0}D_{t}^{\,\alpha}}\,n_{EW; \alpha}(t)$. Hence, Eq. \eqref{15/12-1} can be rewritten as
\begin{equation}\label{15/12-2}
-\tau_{2}^{-\alpha} n_{EW; \alpha}(t) = {_{0}D_{t}^{\,\alpha}}\,n_{EW; \alpha}(t) + \frac{\D}{\D t} n_{EW; \alpha}(t).
\end{equation}
Eq. \eqref{26/10-1} for the EW model is obtained by substituting $k_{EW; \,\alpha}(t)$ into it. That allows one to obtain 
\begin{equation}\label{16/12-1}
\frac{\D}{\D t} \int_{0}^{t} (t - \xi)^{\alpha - 1}\, E_{\alpha, \,\alpha}[-(t-\xi)^{\alpha}/\tau^{\alpha}] \phi_{EW,\,\alpha}(\xi) \D\xi = - \phi_{EW, \, \alpha}(t).
\end{equation}}
\end{exa}

\section{Conclusions}\label{sec5}

We have studied the most popular models of the non-Debye relaxations, namely the HN, JWS, and EW ones. We have shown that using only one assumption concerning the Stieltjes character of the spectral function guarantees the non negativeness of the response and relaxation functions. Such conclusion is not surprising because SFs are the subclass of CMFs and in fact it can be deduced from the survey paper \cite{RGarrappa16}. Our really new results are to pay the readers attention to the fact that we use SFs instead of CMFs and connect the Laplace (L\'{e}vy) exponent $\Psi(s)$ with the integral kernels $M(t)$ and $k(t)$, basic objects which through the evolution equations govern the behavior of the response and relaxation functions. This observation is grace to the fact that $\Psi(s)$ is CBF hence $s/\Psi(s)$ is also CBF. That allows us to join these functions with $M(t)$ and $k(t)$. Namely, $M(t) = [\Psi(s)]^{-1}$ and $k(t) = \Psi(s)/s$. Hence, one stochastic process underlying the relaxation can be described in two-fold - either by $M(t)$ or by $k(t)$. Throughout the paper we were able to reconstruct the previously known form of $M(t)$ for the HN relaxation model \cite{AAKhamzin14, CFAERosa15} and to find $M(t)$ for the JWS and EW models as well as $k(t)$ for all models investigated models. Relevant kernels are itemized in Table \ref{tab2} whereas their shapes in $s$ domain are presented in Table \ref{tab1}. We have also shown that $M(t)$ and $k(t)$ are SFs so they can be called fading memories. Moreover, $M(t)$ and $k(t)$ satisfy the classical Sonine equation and have integrable singularities at zero, thus form the Sonine pairs. We provided three examples of them: $\Big(k_{HN}(t), M_{HN}(t)\Big)$, $\Big(k_{JWS}(t), M_{JWS}(t)\Big)$, and $\Big(k_{EW}(t), M_{EW}(t)\Big)$ which, each other, lead to two equations: the integral and the integro-differential one which are mutually coupled by the Sonine equation for the memories $M(t)$ and $k(t)$. 

A byproduct of our considerations is providing explicit expressions for functions belonging to the Mittag-Leffler family, namely for the first derivative of $E_{\nu, 1}^{\lambda}(x)$ and $E_{(\nu_{1}, \nu_{2}), 1}(x)$. We emphasize that these formulae were derived using physical arguments - relation between the response and relaxation functions.

\vspace{6pt}


\section*{Acknowledgments}

K.G. and A.H. were supported by the Polish National Center for Science (NCN) research grant OPUS12 
no. UMO-2016/23/B/ST3/01714.

\appendix
\section{Properties of SFs and CBFs}

In this Appendix we list the properties of SFs and CBFs which we have extensively used in our construction and proofs. 

\subsection*{Properties of SFs which identify them as subclass of CMFs}

\begin{itemize}

\item[(s1)] The set of SFs is closed under pointwise limits: if $(f_{n})_{n\in\mathbb{N}}$ is a SF and if the limit $\lim_{n\to\infty} f_{n}(s) = f(s)$ exist for all $s > 0$, then $f$ is also a SF, see \cite[Theorem 2.2 (iii)]{RLSchilling12}

\item[(s2)] $f$ is a CBF if, and only if, $(u + f)^{-1}$ is a SF for every $u > 0$, see \cite[Theorem 7.5]{RLSchilling12}

\end{itemize}


\subsection*{Properties of CBFs}

\begin{itemize}
\item[(cb1)]
The set CBF is a convex cone: $s f_{1} + t f_{2}\in {\rm CBF}$ for all $s, t \gg 0$ and $f_{1}, f_{2} \in {\rm CBF}$, see \cite[Corollary 7.6 (i)]{RLSchilling12}.

\item[(cb2)]
According to \cite[Corollary 7.9 (i) and (ii)]{RLSchilling12} $CBF \circ SF \subset SF$ and $SF \circ CBF \subset SF$.

\item[(cb3)]
$f(s)\in CBF$ if, and only if, $1/f(1/s) \in CBF$, see \cite[Eq. (7.1)]{RLSchilling12}.

\item[(cb4)]
\cite[Proposition 7.7]{RLSchilling12} says that if $g\in SF$ is bounded, then $g(0+) - g \in CBF$. Conversely, if the function $f\in CBF$ is bounded, there exist some constant $c > 0$ and some bounded function $g\in SF$, $\lim_{s\in\infty} g(s) = 0$, such that $f = c-g$; then $c = f(0+) + g(0+)$.

\item[(cb5)]
\cite[Proposition 7.1]{RLSchilling12} says that $f(s)$ is in $CBF$, $f \neq 0$, if, and only if $s/f(s)$ is in $CBF$.

\item[(cb6)]
\cite[Theorem 6.2 (ii)]{RLSchilling12} says that if $f(s)$ is a CBF then $f(s)/s$ is a SF.

\item[(cb7)] The set of CBFs is closed under pointwise limits, see \cite[Corollary 7.6 (ii)]{RLSchilling12} 

\item[(cb8)] If $f$ is a CBF then $(u + f)^{-1}$ is a SF for every $u > 0$.
\end{itemize}

\section{Mittag-Leffler function and binomial Mittag-Leffler function}\label{app2}

\subsection{Three parameter Mittag-Leffler function}

The series form of three parameter Mittag-Leffler function $E_{\nu, \mu}^{\lambda}(z)$ \cite{EDeOliveira11, TRPrabharak71, KGorska18} reads
\begin{equation}\label{9/11-1}
E_{\nu, \,\mu}^{\lambda}(z) = \frac{1}{\Gamma(\lambda)}\,\sum_{r\geq 0} \frac{\Gamma(\lambda + r) z^{r}}{r! \Gamma(\mu + \nu r)}
\end{equation}
$z\in\mathbb{C}$ and for the real argument it is involved in the Prabhakar function $t^{\mu - 1} E_{\nu, \mu}^{\lambda}(-a t^{\alpha})$ where $t \geq 0$. The Laplace transform of Prabhakar function reads
\begin{equation}\label{11/11-1a}
\mathcal{L}[t^{\mu - 1} E_{\nu, \,\mu}^{\lambda}(-a t^{\nu}); z] = z^{\nu\lambda - \mu}(a + z^{\nu})^{-\lambda} \quad \text{for} \quad \Re({\mu}), \Re(z) > 0, \,\,\, |z| > |a|^{1/\Re(\nu)}
\end{equation}
\cite{TRPrabharak71} whereas the integral representation of Prabhakar function can be found in \cite[Eq. (11)]{KGorska18} or \cite[Eq. (17)]{KGorska20}
\begin{equation}\label{12/11-1a}
t^{\,\nu\mu - 1} E_{\nu, \,\nu\mu}^{\mu}(-a t^{\nu}) = \frac{1}{\Gamma(\mu)} \int_{0}^{\infty} \E^{-a u} u^{\mu - 1} g_{\nu}(u, t) \D u.
\end{equation}
For $\Re(\nu), \Re(\lambda) > 0$, $\Re(\mu) > n$ and $n \in \mathbb{N}$ the formula (11.5) of \cite{HJHaubold11} holds
\begin{equation}\label{15/11-2b}
\frac{\D^{\,n}}{\D x^{n}} [x^{\mu - 1} E_{\nu, \,\mu}^{\lambda}(a x^{\nu})] = x^{\mu - 1 - n} E_{\nu, \,\mu - n}^{\lambda}(a x^{\nu}).
\end{equation}

\subsection{Binomial Mittag-Leffler function}

The binomial Mittag-Leffler function $E_{(\nu_{1}, \nu_{2}), \mu}(x_{1}, x_{2})$ \cite{RHilfer17, TSandev19} yields
\begin{equation}\label{9/11-2}
E_{(\nu_{1}, \nu_{2}), \mu}(x_{1}, x_{2}) = \sum_{k\geq 0} \mathop{\sum_{l_{1}, l_{2} \geq 0}}_{l_{1} + l_{2} = k} \frac{k!}{l_{1}! l_{2}!}\, \frac{x_{1}^{l_{1}} x_{2}^{l_{2}}}{\Gamma(\mu + \nu_{1} l_{1} + \nu_{2} l_{2})}
\end{equation}
with $x_{1}$ and $x_{2}$ being real. It can be expressed as the series of three parameter Mittag-Leffler functions, namely
\begin{align}\label{23/11-1}
\begin{split}
E_{(\nu_{1}, \nu_{2}), \mu}(x_{1}, x_{2}) & = \sum_{r \geq 0} x_{1}^{r} E_{\nu_{2}, \, \nu_{1} r + \mu}^{1 + r}(x_{2}) \\
& = \sum_{r \geq 0} x_{2}^{r} E_{\nu_{1}, \, \nu_{2} r + \mu}^{1 + r}(x_{1}).
\end{split}
\end{align}
\begin{prf}
Eqs. \eqref{23/11-1} come from Eq. \eqref{9/11-2} by using the restriction $l_{1} + l_{2} = k$. This requirement allows one to change the double sum over $l_{1}$ and $l_{2}$ onto the one sum over $l_{1}$. Thus, Eq. \eqref{9/11-2} can be expressed in the form
\begin{align}\label{21/12-1}
\begin{split}
E_{(\nu_{1}, \nu_{2}), \mu}(x_{1}, x_{2}) & = \sum_{k\geq 0} \sum_{l_{1} = 0}^{k} \binom{k}{l_{1}} \frac{x_{1}^{l_{1}} x_{2}^{k-l_{1}}}{\Gamma[\mu + \nu_{1} l_{1} + \nu_{2}(k - l_{1})]} \\
& = \sum_{l_{1} \geq 0} \sum_{k \geq l_{1}} \binom{k}{l_{1}} \frac{x_{1}^{l_{1}} x_{2}^{k-l_{1}}}{\Gamma[\mu + \nu_{1} l_{1} + \nu_{2}(k - l_{1})]}.
\end{split}
\end{align}
Changing now the summation index $k - l_{1}$ onto $r$ we have
\begin{equation}\label{21/12-2}
E_{(\nu_{1}, \nu_{2}), \mu}(x_{1}, x_{2}) = \sum_{l_{1} \geq 0} \sum_{r \geq 0} \binom{r+l_{1}}{l_{1}} \frac{x_{1}^{l_{1}} x_{2}^{r}}{\Gamma(\mu + \nu_{1} l_{1} + \nu_{2}r)}
\end{equation}
and using the series expression of the three parameter Mittag-Leffler function we can obtain Eqs. \eqref{23/11-1}. \qed
\end{prf}
\noindent
Its Laplace transform can be found in \cite{TSandev19} and it reads
\begin{equation}\label{15/11-1b}
\mathcal{L}[t^{\beta-1} E_{(\alpha_{1}, \alpha_{2}), \beta}(-a_{1} t^{\alpha_{1}}, - a_{2} t^{\alpha_{2}}); t] = \frac{s^{-\beta}}{1 + a_{1} s^{-\alpha_{1}} + a_{2} s^{-\alpha_{2}}}.
\end{equation}


\end{document}